\documentclass[twocolumn,prl,superscriptaddress,longbibliography]{revtex4-1}
\usepackage{graphicx,amsmath,amssymb}
\usepackage[colorlinks=true, citecolor=blue, urlcolor=blue, linkcolor=red ]{hyperref}
\renewcommand{\section}[1]{{\par\it #1.---}\ignorespaces}

\begin{document}
\title{Floquet Topological Phases of Non-Hermitian Systems}
\author{Hong Wu}
\affiliation{School of Physical Science and Technology, Lanzhou University, Lanzhou 730000, China}
\author{Jun-Hong An}
\email{anjhong@lzu.edu.cn}
\affiliation{School of Physical Science and Technology, Lanzhou University, Lanzhou 730000, China}
\begin{abstract}
The non-Hermiticity caused breakdown of the bulk-boundary correspondence (BBC) in topological phase transition was cured by the skin effect for the systems with chiral symmetry and translation invariance. However, periodic driving, as an active tool in engineering exotic topological phases, breaks the chiral symmetry, and the inevitable disorder destroys the translation invariance. Here, we propose a scheme to retrieve the BBC and establish a complete description of the topological phases of the periodically driven non-Hermitian system both with and without the translation invariance. The demonstration of our method in the non-Hermitian Su-Schrieffer-Heeger model shows that exotic non-Hermitian topological phases of widely tunable numbers of edge states and Floquet topological Anderson insulator are induced by the periodic driving and the disorder. Our result supplies a useful way to artificially synthesize exotic phases by periodic driving in the non-Hermitian system.

\end{abstract}
\maketitle

\section{Introduction}
Topological phases in non-Hermitian systems have attracted much attention both theoretically \cite{PhysRevLett.116.133903, PhysRevLett.118.040401,PhysRevLett.121.026808,PhysRevLett.121.086803,PhysRevLett.121.136802,PhysRevB.99.081103,PhysRevLett.120.146402,PhysRevLett.123.066404,PhysRevX.8.031079,PhysRevX.9.041015,PhysRevLett.124.056802,PhysRevLett.124.086801,PhysRevB.99.201103,PhysRevB.101.020201,PhysRevLett.123.206404,PhysRevLett.120.065301,PhysRevB.99.125155,PhysRevLett.122.237601,PhysRevB.100.035102,PhysRevLett.124.073603,PhysRevLett.122.076801,PhysRevLett.123.016805,PhysRevLett.122.195501,PhysRevB.100.081401,PhysRevA.99.032121, PhysRevB.100.081104,PhysRevB.98.245130,PhysRevB.99.075130,PhysRevB.100.245116,zhang2020bulkboundary,PhysRevLett.121.026808,PhysRevA.99.052118,zhang2019nonhermitian,wang2019defective} and experimentally \cite{PhysRevLett.115.040402,PhysRevX.6.021007,Xiao2017,PhysRevLett.121.124501,PhysRevLett.123.165701,Zhao1163,ghatak2019observation,xiao2019observation}. Many interesting characters have been found in different non-Hermitian systems \cite{PhysRevLett.120.065301,PhysRevB.99.125155,PhysRevLett.122.237601,PhysRevB.100.035102,PhysRevLett.124.073603,PhysRevLett.122.076801,PhysRevLett.123.016805,PhysRevLett.122.195501,PhysRevB.100.081401,PhysRevA.99.032121,PhysRevB.100.081104,PhysRevB.98.245130,PhysRevB.99.075130,PhysRevB.100.245116,zhang2020bulkboundary}. One of their unique features is that not only the edge states but also the nontopologically protected bulk states are localized at the edges, which is called skin effect \cite{PhysRevX.9.041015,PhysRevLett.124.056802,PhysRevB.99.201103,PhysRevB.101.020201,PhysRevLett.124.086801,PhysRevB.97.121401,Xiong_2018,zhang2019correspondence,yang2019auxiliary}. It causes that one cannot characterize the edge states by the topological properties of the bulk spectrum. This is the non-Hermiticity induced breakdown of bulk-boundary correspondence (BBC) \cite{PhysRevLett.116.133903, PhysRevLett.118.040401,PhysRevLett.121.026808,PhysRevLett.121.086803,PhysRevLett.121.136802}, which lays the
foundation for the classification of topological phases in Hermitian systems \cite{RevModPhys.82.3045,RevModPhys.83.1057,RevModPhys.88.021004,RevModPhys.88.035005}. To describe the topological features of the edge states, many strategies including biorthogonal eigenstate \cite{PhysRevLett.121.026808}, singular-value decomposition \cite{PhysRevA.99.052118}, gauge transformation \cite{PhysRevB.100.054105}, and modified periodic boundary condition \cite{PhysRevB.100.165430} have been proposed. A milestone among these is the non-Bloch band theory established in the generalized Brillouin zone (BZ) for the one-dimensional (1D) chirally symmetric and translation invariant systems \cite{PhysRevLett.121.086803,PhysRevLett.123.066404}, which is recently generalized to the system without chiral symmetry \cite{zhang2019correspondence,yang2019auxiliary}.

Coherent control via periodic driving dubbed as Floquet engineering has become a versatile tool in artificially synthesizing exotic topological phases in systems of ultracold atoms \cite{RevModPhys.89.011004,PhysRevLett.116.205301}, photonics \cite{Rechtsman2013,PhysRevLett.122.173901}, superconductor qubits \cite{Roushan2017}, and graphene \cite{McIver2020}. Parallel to the topological phases in static systems, the topological phases in periodically driven systems are called Floquet topological phases (FTPs). Many intriguing FTPs absent in static systems \cite{PhysRevLett.111.136402,PhysRevLett.123.066403,PhysRevLett.118.115301,PhysRevLett.124.057001,PhysRevX.3.031005,PhysRevB.88.121406,PhysRevA.93.043618,PhysRevLett.120.243602,PhysRevLett.121.036402,PhysRevB.99.214303,PhysRevA.98.063847,PhysRevB.97.245430} have been simulated by periodic driving in Hermitian systems. The key role played by periodic driving is changing symmetry and inducing an effective long-range hopping in lattice systems \cite{PhysRevB.87.201109,PhysRevB.93.184306,PhysRevA.100.023622}. A natural question is what controllable topological characters can periodic driving bring to non-Hermitian systems. Given the fact that the chiral symmetry can be broken by periodic driving, one cannot apply the well developed non-Bloch band theory of 1D chirally symmetric static systems \cite{PhysRevLett.121.086803} to the periodic ones for recovering the BBC and defining topological invariants. Without touching the topological characterization, the transport phenomena of the non-Hermitian Floquet edge states was studied in Refs. \cite{PhysRevLett.123.190403,PhysRevB.100.045423}. For some special cases in the absence of the skin effect, the topological numbers were defined in the traditional BZ \cite{PhysRevB.98.205417,PhysRevB.101.014306}. Recent study reveals that the BBC is approximately recoverable only for small intercell coupling \cite{PhysRevB.101.045415}. Further, the inevitable spatial disorder also invalidates the non-Bloch band theory to restore the BBC. Thus, a general theory to characterize the non-Hermitian FTPs is still lacking.

In this work, we investigate the FTPs in the periodically driven non-Hermitian systems. A general description is established to characterize the FTPs of such nonequilibrium systems both in the momentum and the real spaces. The main idea to characterize the FTPs in both of the spaces is to restore the chiral symmetry of the periodically driven systems by the proposed similarity transformations, which keep the quasienergy spectrum unchanged. Taking the non-Hermitian Su-Schrieffer-Heeger (SSH) model as an example, we find that rich topological phases absent in the static case are created by the periodic driving. The studies on the real-space topological physics in the presence of disorder reveal that the extra phases called non-Hermitian Floquet topological Anderson insulator phases are induced by the disorder. Our results demonstrate that the periodic driving and its constructive interplay with the disorder supply us useful ways to engineer exotic topological phases in the non-Hermitian systems.

\section{Floquet topological phases}
A time-periodic system $ {H}(t)= {H}(t+T)$ with period $T$ has a complete set of basis $|u_\alpha(t)\rangle$ determined by Floquet equation $[ {H}(t)-i\partial_t]|u_\alpha(t)\rangle=\varepsilon_\alpha|u_\alpha(t)\rangle$ such that any state evolves as $|\Psi(t)\rangle=\sum_\alpha c_\alpha e^{-i\varepsilon_\alpha t}|u_\alpha(t)\rangle$ \cite{PhysRevA.7.2203,PhysRevA.91.052122}. Acting as stationary states and eigenenergies of static systems, $|u_\alpha(t)\rangle$ and $\varepsilon_\alpha$ are called quasistationary states and quasienergies, respectively. Being equivalent to ${U}_T|u_\alpha(0)\rangle=e^{-i\varepsilon_\alpha T}|u_\alpha(0)\rangle$ with $U_T$ the one-period evolution operator, the Floquet equation defines an effective Hamiltonian ${H}_\text{eff}=\frac{i}{T}\ln U_T$ whose eigenvalues are the quasienergies. The FTPs are defined in the quasienergy spectrum. Different from the static case, they can occur at both of the quasienergies $0$ and $\pi/T$ \cite{PhysRevB.87.201109}.

Chiral symmetry plays an important role in characterizing the non-Hermitian topological phases \cite{PhysRevLett.118.040401,PhysRevLett.121.026808,PhysRevLett.121.086803,PhysRevB.99.081103,PhysRevLett.123.066404}. However, it cannot be preserved if a periodic driving is applied. Consider a non-Hermitian two-band system $H$ with its parameters periodically driven between two specific $H_1$ and $H_2$ in the respective time duration $T_1$ and $T_2$. Applying the Floquet theorem, we obtain $H_\text{eff}$ from $U_T=e^{-iH_2T_2}e^{-iH_1T_1}$. It can be seen that even $H_j$ ($j=1,2$) have chiral symmetry $SH_jS^{-1}=-H_j$ with $S$ being the chiral operator, $H_\text{eff}$ breaks the symmetry due to $[H_1,H_2]\neq0$.  The absence of the chiral symmetry in $H_\text{eff}$ makes it hard to define the FTPs in a non-Hermitian system by the non-Bloch band theory, which is developed for the chirally symmetric static system \cite{PhysRevLett.121.086803}. We propose the following scheme to resolve this problem. Two similarity transformations $G_j=e^{i(-1)^jH_jT_j/2}$ covert $U_T$ into $\tilde{U}_{T,1}=e^{-iH_1T_1/2}e^{-iH_2T_2}e^{-iH_1T_1/2}$ and $\tilde{U}_{T,2}=e^{-iH_2T_2/2}e^{-iH_1T_1}e^{-iH_2T_2/2}$, from which the defined $\tilde{H}_{\text{eff},j}={i\over T}\ln \tilde{U}_{T,j}$ share the same quasienergies with $H_\text{eff}$ while recover the chiral symmetry of $H_j$. It can be equivalently understood to define new chiral operators $G_j^{-1}S^{-1}G_j$ such that $H_\text{eff}$ obeys the chiral symmetry. The similar scheme was used in Hermitian systems \cite{PhysRevB.90.125143}. As we will see later, the recovered chiral symmetry is significant to characterize the FTPs in the non-Hermitian system both for the translation-invariant and variant cases.

\section{Translation-invariant non-Hermitian system}
If the system is further translation invariant, we can develop a general characterization to the FTPs in the momentum space. The coefficient matrices of $H_j$ are written in the momentum space as $\mathcal{H}_j(k)=\mathbf{h}_j(k)\cdot\pmb{\sigma}$ with $\pmb{\sigma}$ being the Pauli matrices. We readily obtain $\mathcal{H}_\text{eff}({ k})=\mathbf{h}_\text{eff}(k)\cdot{\pmb\sigma}=i\ln[e^{-i\mathcal{H}_2(k)T_2}e^{-i\mathcal{H}_1(k)T_1}]/T$ with the Bloch vector $\mathbf{h}_\text{eff}(k)=-\arccos (\epsilon) \underline{\mathbf{r}}/T$ and
\begin{eqnarray}
\epsilon&=&\cos(T_1 E_1)\cos(T_2 E_2)-\underline{\mathbf{h}}_1\cdot\underline{\mathbf{h}}_2\sin(T_1{E}_1)\sin(T_2E_2),~~~\label{epsl}\\
\mathbf{r}&=& \underline{\mathbf{h}}_1\times\underline{\mathbf{h}}_2\sin(T_1E_1)\sin(T_2E_2)-\underline{\mathbf{h}}_2\cos(T_1E_1)\nonumber\\ &&\times\sin(T_2E_2)-\underline{\mathbf{h}}_1\cos(T_2{E}_2)\sin(T_1{E}_1),\label{varepsilon}
\end{eqnarray}
where $T=T_1+T_2$, $\underline{\mathbf{h}}_j=\mathbf{h}_j/E_j$, and $E_j=\sqrt{\mathbf{h}_j\cdot\mathbf{h}_j}$ is the complex eigen energies of $\mathcal{H}_j(k)$ \cite{SMP}. The FTP transition is associated with the closing of the quasienergy bands, which occurs at the exceptional points for the $k$ and driving parameters satisfying
\begin{eqnarray}
&&T_jE_j=n_j\pi,~n_j\in \mathbb{Z}, \label{gen}\\
&&\text{or}~
\begin{cases}
\underline{\mathbf{h}}_1\cdot\underline{\mathbf{h}}_2=\pm1\\
T_1{E}_1\pm T_2{E}_2=n\pi,~n\in\mathbb{Z}
\end{cases}\label{hh1}
\end{eqnarray}
at the quasienergy zero (or $\pi/T$) if $n$ is even (or odd) \cite{SMP}.
As the condition for the phase transition, Eqs. \eqref{gen} and \eqref{hh1} supply a guideline to manipulate the exceptional points via the driving parameters for engineering various non-Hermitian FTPs at will. They reduce to the results in the Hermitian case \cite{PhysRevB.93.184306,PhysRevA.100.023622} as a special case when the non-Hermitian terms in ${\bf h}_j$ vanish.

We see from Eq. \eqref{varepsilon} that $\mathbf{h}_\text{eff}(k)$ generally has three components even though the chirally symmetric $\mathbf{h}_j$ have only two. It proves that the chiral symmetry is broken by the periodic driving \cite{SMP}. Thanks to the similarity transformation $G_j$, we obtain $\tilde{\mathcal{H}}_{\text{eff},j}(k)$ preserving the chiral symmetry of $\mathcal{H}_j(k)$. Then we can restore the BBC and define proper topological invariants in our periodically driven non-Hermitian system by introducing the generalized BZ in the similar manner as the static system \cite{PhysRevLett.121.086803}. The topological properties of the periodic non-Hermitian system are fully characterized by the two winding numbers $\mathcal{W}_j$ defined in the generalized BZ associated with $\tilde{\mathcal{H}}_{\text{eff},j}$. The number of $0$- and $\pi/T$-mode edge states relates to $\mathcal{W}_j$ as \cite{xiao2019observation, PhysRevB.98.205417}
\begin{equation}
N_0=|\mathcal{W}_1+\mathcal{W}_2|/2,~N_{\pi/T}=|\mathcal{W}_1-\mathcal{W}_2|/2. \label{tpnb}
\end{equation}

Without loss of  generality, we demonstrate our method by the 1D non-Hermitian SSH model \cite{PhysRevLett.42.1698,PhysRevA.89.062102,PhysRevA.97.052115}
\begin{eqnarray}
H=\sum_{l=1}^L[(t_1+\frac{\gamma}{2})a^{\dag}_lb_l+(t_1-\frac{\gamma}{2})b^{\dag}_la_l+t_2(a^{\dag}_lb_{l-1}+\text{h.c.})],~~~\label{Hmat}
\end{eqnarray}where $a_l$ ($b_l$) are the annihilation operators on the sublattice $A$ ($B$) of the $l$th lattice, and $L$ is lattice length. In momentum space and the operator basis $(\tilde{a}_k,\tilde{b}_k)^T$ with $\tilde{a}_k$ ($\tilde{b}_k$) being the Fourier transform of $a_l$ ($b_l$), it reads
\begin{equation}
\mathcal{H}(k)=d_x\sigma_x+(d_y+i{\gamma}/{2})\sigma_y,\label{Hmt}
\end{equation}
where $d_x=t_1+t_2\cos k$ and $d_y=t_2\sin k$. The bands close at $k=\pi$ (or $0$) when $t_1=t_2\pm\gamma/2$ (or $-t_2\pm\gamma/2$). It is in conflict with the result under the open-boundary condition, where the bands close when $t_1=\sqrt{t_2^2+\gamma^2/4}$. It is called the non-Hermiticity caused breakdown of BBC \cite{PhysRevLett.116.133903, PhysRevLett.118.040401,PhysRevLett.121.026808,PhysRevLett.121.086803,PhysRevLett.121.136802}. The problem for the static system with chiral symmetry $\sigma^{-1}_z\mathcal{H}(k)\sigma_z=-\mathcal{H}(k)$ \cite{RevModPhys.88.035005} is cured by the skin effect. Via replacing $e^{ik}$ by $\beta=\sqrt{\lvert \frac{t_1-\gamma/2}{t_1+\gamma/2}\lvert}e^{ik}$, Eq. \eqref{Hmt} is converted into $\mathcal{H}(\beta)=\sum_{n=\pm}R_{n}(\beta)\sigma_{n}$ with $\sigma_{\pm}=(\sigma_x\pm i\sigma_y)/2$ and $R_{\pm}(\beta)=t_1\pm\frac{\gamma}{2}+\beta^{\mp1} t_2$. Here $\beta$ defines a generalized BZ. Its topological property is described by the winding number
$\mathcal{W}=-(\mathcal{W}_{+}-\mathcal{W}_{-})/2$, where $\mathcal{W}_{\pm}=\frac{1}{2\pi}[\arg R_{\pm}(\beta)]_{C_{\beta}}$ with $[\arg R_{\pm}(\beta)]_{C_{\beta}}$ are the phase change of $R_{\pm}$ as $\beta$ counterclockwisely goes along the generalized BZ $C_{\beta}$ \cite{PhysRevLett.121.086803,PhysRevLett.123.066404}. When $|t_1|<\sqrt{t^2_2+\gamma^2/4}$, $\mathcal{W}=1$ and a pair of edge states is formed.

\begin{figure}[tbp]\centering
\includegraphics[width=1\columnwidth]{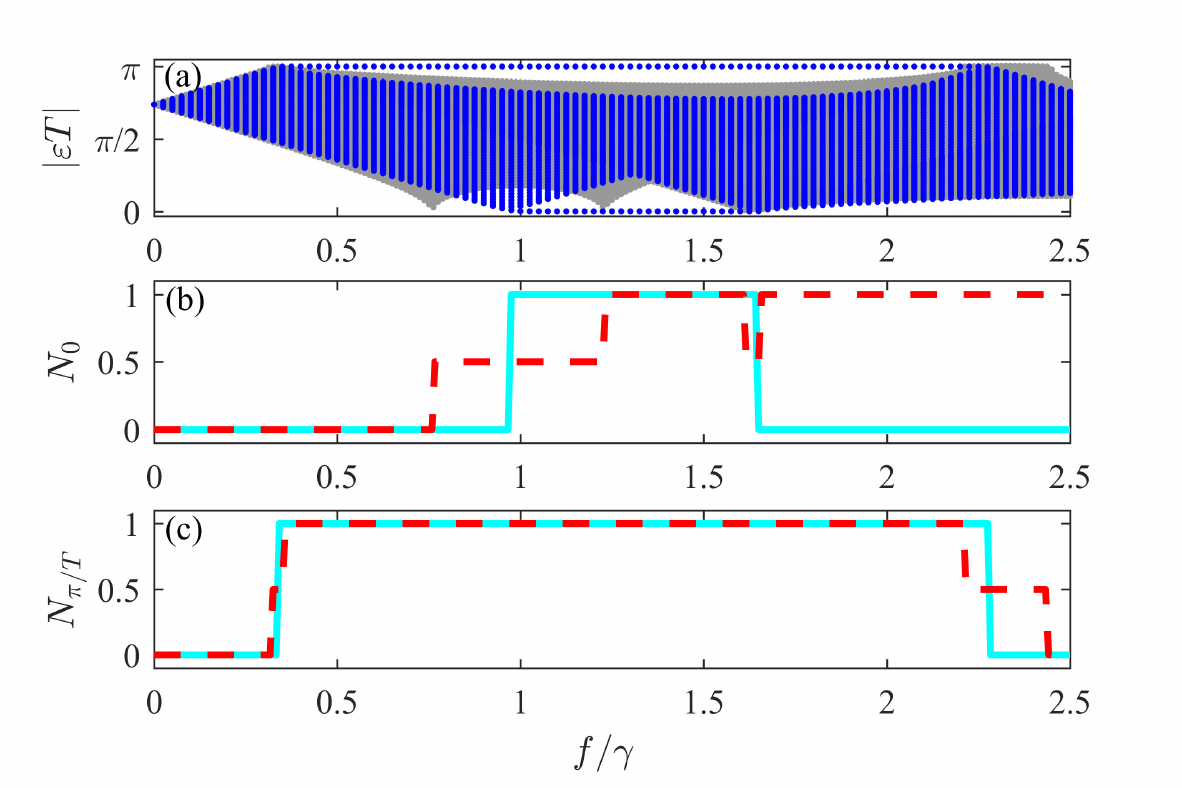}
\caption{(a) Quasienergy spectra with the change of the driving amplitude under the open (blue lines) and periodic (gray lines) boundary conditions. Numbers of $0$-mode (b) and $\pi/T$-mode (c) edge states defined in the conventional (red dashed) and generalized (cyan solid) BZ. We use $t_1=2.0\gamma$, $T_1=T_2=0.6\gamma^{-1}$, $q=3.0$, and $L=80$.}\label{engsp}
\end{figure}

Choosing the periodic driving as
\begin{equation}
t_2(t) =\begin{cases}f ,& t\in\lbrack mT, mT+T_1)\\q\,f,& t\in\lbrack mT+T_1, (m+1)T), \end{cases}~m\in \mathbb{Z}, \label{procotol}
\end{equation}
we now investigate the FTPs in our periodically driven non-Hermitian SSH model. Figure \ref{engsp}(a) shows the quasienergy spectrum under the open-boundary condition. It indicates that even the static system when $f=0$ is topologically trivial, diverse topological phases at the quasienergies $0$ and $\pi/T$ can be created by the periodic driving. However, this quasienergy spectrum has a dramatic difference from the one under the periodic-boundary condition, which takes $\sqrt{\mathbf{h}_\text{eff}(k)\cdot\mathbf{h}_\text{eff}(k)}$. It reveals that the non-Hermiticity induced breakdown of BBC occurs in our periodically driven system too. To solve this problem, we introduce the generalized BZ via replacing $e^{ik}$ in $\mathcal{H}_\text{eff}(k)$ by $\beta$ \cite{SMP}. Then the effective Hamiltonian is converted to $\mathcal{H}_\text{eff}(\beta)$. First, $\mathcal{H}_\text{eff}(\beta)$ correctly explains the exceptional points of the quasienergies under the open-boundary condition. Remembering ${\bf h}(t)=[t_1+t_2(t)(\beta+\beta^{-1})/2,i[\gamma+t_2(t)(\beta^{-1}-\beta)]/2,0]$ and using Eqs. \eqref{gen} and \eqref{hh1} by setting $t_1>\gamma/2>0$, we obtain the phase-transition conditions as follows.

\textbf{Case I:} $\underline{{\bf h}}_1\cdot\underline{{\bf h}}_2=1$. We can check that Eqs. \eqref{hh1} induce
\begin{eqnarray}
T_1|\kappa+e^{i\alpha}f|+T_2|\kappa+e^{i\alpha}qf|=n_\alpha\pi,~(n_\alpha\in \mathbb{Z}) \label{zercon}
\end{eqnarray}for $k$ in $\beta$ being $\alpha=0$ or $\pi$, where $\kappa=\sqrt{t_1^2-\gamma^2/4}$. Here $\text{sgn}[(\kappa-f)(\kappa-qf)]=1$ is further needed for $\alpha=\pi$.

\textbf{Case II:} $\underline{{\bf h}}_1\cdot\underline{{\bf h}}_2=-1$ requires $k=\pi$ when $\text{sgn}[(\kappa-f)(\kappa-qf)]=-1$. Then Eqs. \eqref{hh1} give
\begin{eqnarray}
T_1|\kappa-f|-T_2|\kappa-qf|=n_\pi\pi. \label{negcon}
\end{eqnarray}

\textbf{Case III:} According to Eq. \eqref{gen}, any $k$ in $\beta$ satisfying
\begin{eqnarray}
T_1E_1=n_1\pi,~T_2E_2=n_2\pi,~(n_1,~n_2\in \mathbb{Z})  \label{finaa}
\end{eqnarray}
contributes to the band closing.

Taking care of the skin effect via introducing $\beta$, Eqs. \eqref{zercon}-\eqref{finaa} perfectly describe the band closing of the quasienergy spectrum under the open-boundary condition. The $\pi/T$-mode band-closing points at $f\simeq0.34\gamma$ and $2.27\gamma$ in Fig. \ref{engsp}(a) are obtainable from Eqs. \eqref{zercon} with $n_0=n_\pi=1$. The $0$-mode ones at $f\simeq 0.97\gamma$ and $1.65\gamma$ are obtainable from Eqs. \eqref{negcon} with $n_\pi=0$ and \eqref{zercon} with $n_0=2$, respectively. Thus the BBC has been successfully retrieved in our periodically driven system.

\begin{figure}[tbp]
\centering
\includegraphics[width=1\columnwidth]{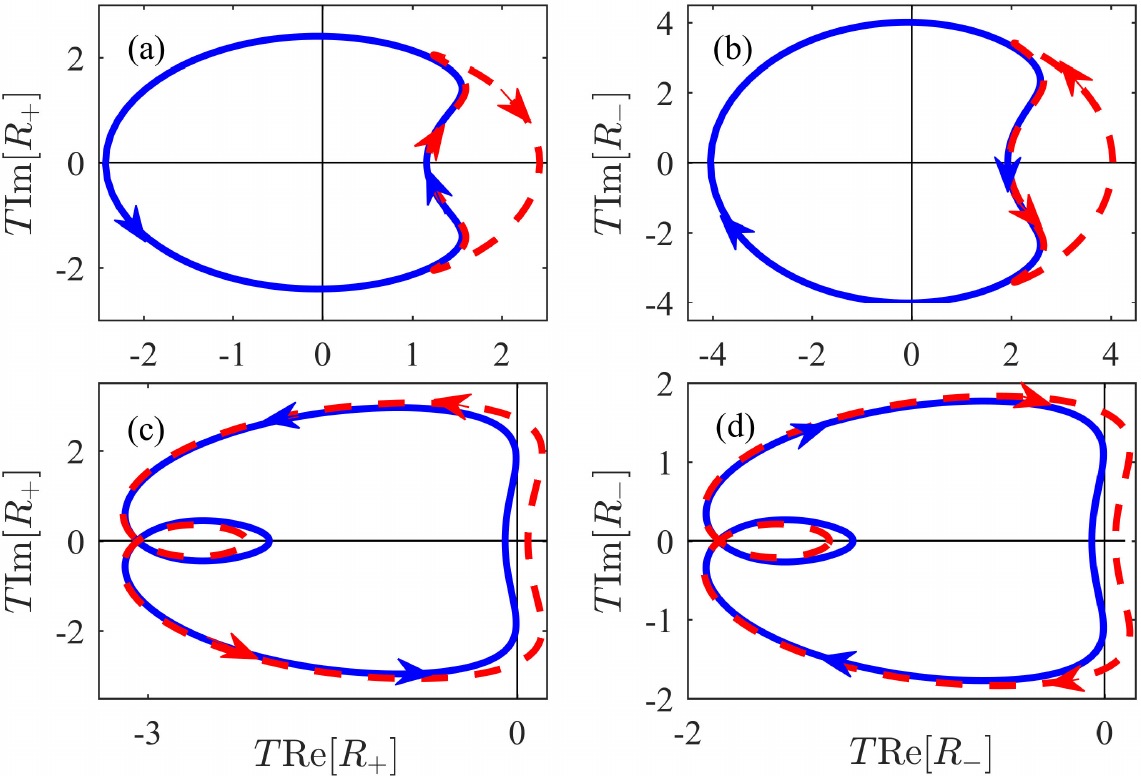}
\caption{Trajectories of $R_\pm$ in $\mathcal{\tilde H}_\text{eff,1}(\beta)$ with $k$ in $\beta$ running from $0$ to $2\pi$ when $f$ crosses the phase boundaries. The winding number $\mathcal{W}_1$ changes from $0$ when $f=0.335\gamma$ (red dashed) to 1 when $0.345\gamma$ (blue solid) in (a) and (b); and from 1 when $f=0.94\gamma$ (red dashed) to 0 when $1.0\gamma$ (blue solid) in (c) and (d). Others parameters are the same as Fig. \ref{engsp}.} \label{traj}
\end{figure}
Second, the FTPs of the quasienergy spectrum under the open boundary condition are well characterized by the two winding numbers $\mathcal{W}_j$ defined in $\tilde{\mathcal H}_{\text{eff},j}$. According to Eq. \eqref{tpnb}, we plot in Figs. \ref{engsp}(b) and \ref{engsp}(c) the numbers of $0$-mode and $\pi/T$-mode edge states calculated from the conventional and generalized BZs. Although qualitatively capturing the  exceptional points of the quasienergy under the periodic-boundary condition, the ill-defined topological numbers from the conventional BZ nonphysically take half integers. However, the ones from the generalized BZ correctly count the number of the edge states. It is called the non-Bloch BBC \cite{PhysRevLett.121.086803,PhysRevLett.123.066404}. Note that, absent in the static system, such correspondence for the $\pi/T$-mode edge states is unique in our periodic system.

Third, the topological change of the quasienergy spectrum is reflected by $\mathcal{H}_\text{eff}(\beta)$. We plot in Fig. \ref{traj} the trajectories of $R_\pm$ in $\mathcal{\tilde H}_\text{eff,1}(\beta)$ when $f$ increases across the phase borders. Figures \ref{traj}(a) and \ref{traj}(b) show that $R_\pm$ have no wrapping to the origin and thus $\mathcal{W}_1=0$ before the $\pi/T$-mode phase transition. When $f$ increases across the critical point, $R_\pm$ at the neighbourhood of $k=0$ changes such that $\varepsilon=\sqrt{R_+R_-}$ crosses $\pi/T$. Due to its periodicity, $\varepsilon$ abruptly jumps to $-\pi/T$ keeping the direction of $R_\pm$ unchanged. Then an anticlockwise and a clockwise wrappings to the origin are formed by $R_+$ and $R_-$, respectively, and thus $\mathcal{W}_1=1$. Figures \ref{traj}(c) and \ref{traj}(d) show that $\mathcal{W}_1$ changes from 1 to 0, where $R_\pm$ at the neighbourhood of $k=\pi$ changes such that $\varepsilon$ crosses the quasienergy 0. This gives a geometric picture to the FTP transition in Fig. \ref{engsp}.

\begin{figure}[tbp]
\centering
\includegraphics[width=1\columnwidth]{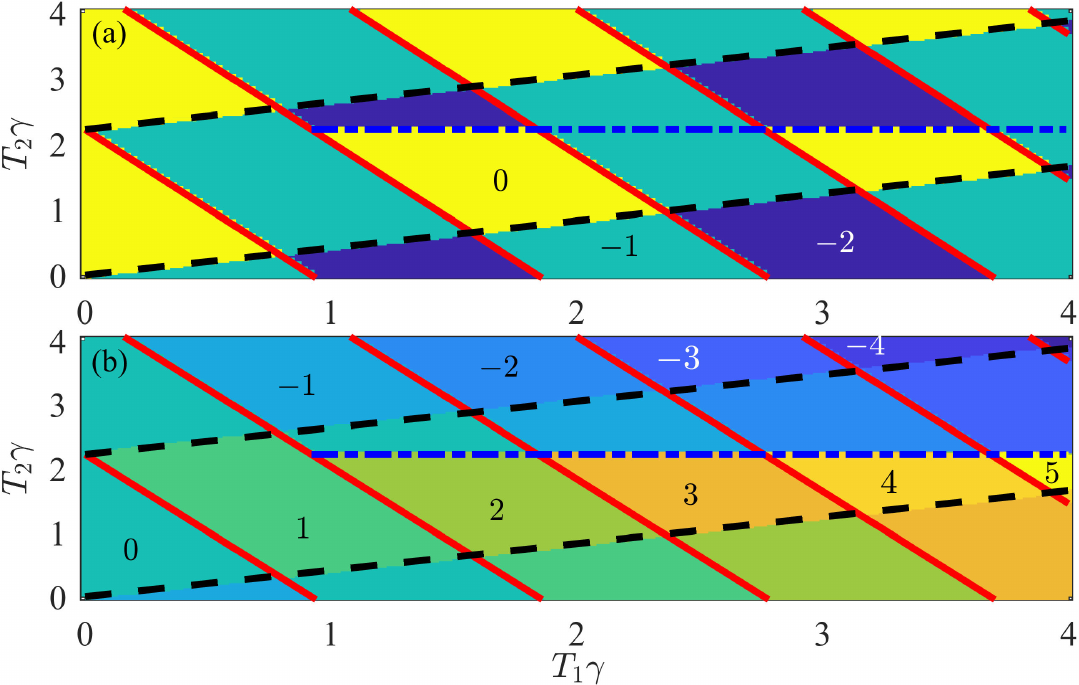}
\caption{Phase diagram characterized by $\mathcal{W}_1$ (a) and $\mathcal{W}_2$ (b). The red solid, the black dashed, and the blue dot-dashed lines are the phase boundaries from Eqs. \eqref{zercon}, \eqref{negcon}, and \eqref{finaa}, respectively. We use $t_1=1.5\gamma$, $f=2\gamma$, and $q=0$.} \label{pad}
\end{figure}

As a useful tool in controlling the exceptional points, the periodic driving enables us to realize not only the topological phases inaccessible in the same static-system condition, but also rich phases completely absent in its original static system. Figure \ref{pad} shows the phase diagram in the $T_1$-$T_2$ plane. A widely tunable number of $\mathcal{W}_j$ and edge states are induced by changing the driving parameters. The presence of such rich phases originates from the distinguished role of periodic driving in simulating an effective long-range hopping in different lattices \cite{PhysRevB.87.201109,PhysRevB.93.184306,PhysRevA.100.023622}. The phase boundaries in red solid lines (black dashed lines) are perfectly described by Eq. \eqref{zercon} with $\alpha=0$ [by Eq. \eqref{negcon}]. $T_2E_2=\pi$ in Eqs. \eqref{finaa} is satisfied by $T_2\simeq 2.22/\gamma$. $T_1E_1=n_1\pi$ is satisfied by $T_1\simeq n_1\pi/(\gamma\sqrt{6+5.66\cos k})$. When $k$ runs from $0$ to $\pi$ for given $n_1$, a series line segments with a common $T_2\simeq 2.22/\gamma$ (see the blue dot-dashed line in Fig. \ref{pad}) are formed, which all give the phase boundaries. We see that our analytical method successfully describes the FTPs in the periodically driven non-Hermitian system. The result reveals that, without changing the intrinsic parameters in the static system, the periodic driving supplies us another control dimension to adjust the numbers of the non-Hermitian topological edge states. This is useful in the application of non-Hermitian topological physics.

\begin{figure}[tbp]
\centering
\includegraphics[width=1\columnwidth]{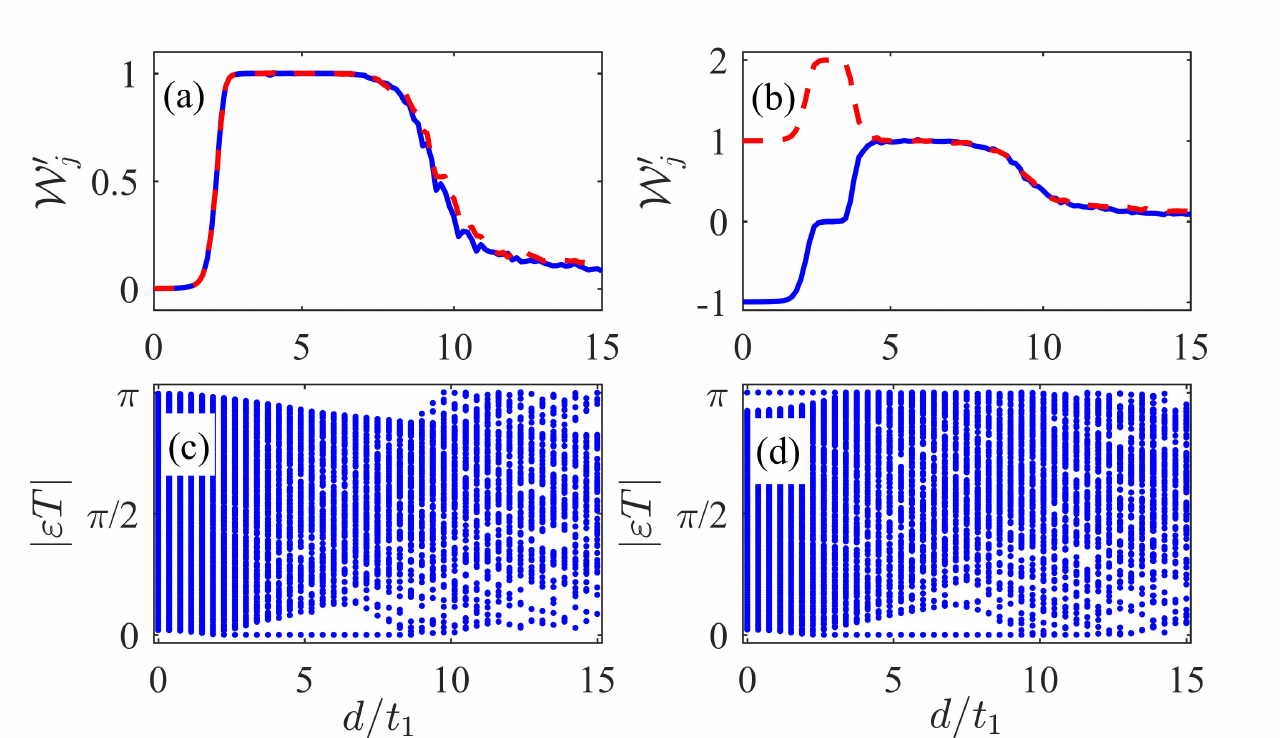}
\caption{Real-space winding numbers $\mathcal{W}_1'$ (blue solid) and $\mathcal{W}_2'$ (red dashed) in (a), (b) and the corresponding quasienergies in (c), (d) with the change of the disorder strength. $T_1=T_2=0.8t_1^{-1}$ in (a), (c) and $0.9t_1^{-1}$ in (b), (d). Other parameters are $\gamma=0.2t_1$, $f=0.48t_1$, $q=3.0$, $L=160$, and $\ell=40$. (a), (b) is obtained after 250 times average to the disorder. } \label{dsord}
\end{figure}
\section{Translation-variant non-Hermitian system} When the translation invariance of Eq. \eqref{Hmat} is broken by the disorder $d\xi_l$ in the non-Hermitian term $\gamma$, where $\xi_l\in[-0.5,0.5]$ is the disorder in the $l$th cell with strength $d$, we cannot work in the momentum space anymore. The non-Bloch band theory is inapplicable too. However, we still may characterize the FTPs by the chirally symmetric $\tilde{\mathcal H}_{\text{eff},j}$ in the real space. Regarding $l\in[1,\ell]$ and $[L-\ell+1,L]$ of the chain as the boundaries, we define the real-space winding numbers \cite{PhysRevLett.123.246801}
\begin{equation}
\mathcal{W}_j'=\frac{1}{2L'}\text{Tr}'(SQ_j[Q_j,X]).
\end{equation}Here $S_{ls,l's'}=\delta_{ll'}(\sigma_z)_{ss'}$ is the chiral operator and $X_{ls,l's'}=l\delta_{ll'}\delta_{ss'}$ with $s,s'=A,B$ being the sublattices, $Q_j=\sum'_n(|n^\text{R}_j\rangle\langle n^\text{L}_j|-S|n^\text{R}_j\rangle\langle n^\text{L}_j|S^\dag)$ with $\tilde{\mathcal H}_{\text{eff},j}|n_j^\text{R}\rangle=\varepsilon_{j,n}|n_j^\text{R}\rangle
$ and $\tilde{\mathcal H}^\dag_{\text{eff},j}|n_j^\text{L}\rangle=\varepsilon^*_{j,n}|n_j^\text{L}\rangle$, and $\text{Tr}'$ denotes the trace over the middle interval with length $L'=L-2\ell$ and $\sum_n'$ denotes the summation to the bulk states. We can check that $\mathcal{W}_j'$ return to $\mathcal{W}$ when $d=0$. Thus, more general than $\mathcal{W}_j$, the real-space $\mathcal{W}_j'$ can give a unified description to the FTPs of the non-Hermitian system both for the translation-invariant and variant cases.

Figure \ref{dsord} shows the winding numbers $\mathcal{W}_j'$ and the quasienergies with the change of the disorder strength. We can see from Figs. \ref{dsord}(a) and \ref{dsord}(c) that the topological trivial character of the disorder-free case is robust when the disorder is weak for $d\lesssim 2t_1$. With the increase of $d$, it is remarkable to find that a $0$-mode edge state is triggered in a wide range $d\in (2,10)t_1$. The disorder-induced edge state has been found in static Hermitian systems \cite{PhysRevLett.102.136806,Meier929,PhysRevLett.105.216601,PhysRevX.6.021013}. Analogous to that, we call the similar state occurred in our periodically driven non-Hermitian system as Floquet topological Anderson insulator phase. Its presence can be further confirmed by Figs. \ref{dsord}(b) and \ref{dsord}(d), where a $\pi/T$-mode edge state exists in the disorder-free case. Here, it is interesting to observe a coexisted regime of the $\pi/T$-mode edge state and 0-mode Floquet topological Anderson insulator state. Both of the states are absent in the static system. However, in the strong disorder regime, the bands close and all the edge states disappear, which is compatible to the result in the Hermitian case \cite{PhysRevLett.121.126803}.

\section{Discussions and conclusion} Note the chiral symmetry is not recoverable in some driving cases \cite{Zhang2017}, where the non-Hermitian FTPs can be described by a $Z_2$ topological invariant \cite{SMP}. Our result is realizable in the present experimental state of art of photonics, where the non-Hermitian topological phases of the SSH model \cite{PhysRevLett.115.040402,PhysRevResearch.2.013280} and the Hermitian FTPs \cite{Rechtsman2013,PhysRevLett.122.173901} have been observed.

We have investigated the topological phases in periodically driven non-Hermitian systems. A scheme is proposed to retrieve the BBC, based on which a complete description to the FTPs is established for such non-Hermitian systems both with and without the translation invariance. Taking the SSH model as an example, we have found that diverse exotic FTPs can be created from the topologically trivial static system by the periodic driving. Further study reveals that the Floquet topological Anderson topological insulator phases can be triggered by the moderate-strength disorder. Exhibiting a wide perspective of controllability, our results hopefully promote further studies of both fundamental physics and potential applications of rich non-Hermitian FTPs.

\section{Acknowledgments}
The work is supported by the National Natural Science Foundation (Grant Nos. 11875150 and 11834005) and the Fundamental Research Funds for the Central Universities of China.

\bibliography{references.bib}

\end{document}

\title{Supplemental material for ``Floquet Topological Phases of Non-Hermitian Systems''}
\author{Hong Wu}
\affiliation{School of Physical Science and Technology, Lanzhou University, Lanzhou 730000, China}
\author{Jun-Hong An}
\email{anjhong@lzu.edu.cn}
\affiliation{School of Physical Science and Technology, Lanzhou University, Lanzhou 730000, China}

\maketitle

\section{Derivation of Eqs. (1)-(4)}
According to the Euler's formula of the Pauli matrices
\begin{equation}
\begin{split}
e^{-i\mathbf{h}_j\cdot\pmb{\sigma}T_j}&=\cos(E_jT_j)-i\sin(E_jT_j)\underline{\mathbf{h}}_j\cdot\pmb{\sigma}
\end{split}
\end{equation}
with $E_j=\sqrt{\mathbf{h}_j\cdot\mathbf{h}_j}$ and $\underline{\mathbf{h}}_j=\mathbf{h}_j/E_j$, we have
\begin{eqnarray}
U(T)&=&e^{-i\mathbf{h}_2\cdot\pmb{\sigma} T_2}e^{-i\mathbf{h}_1\cdot\pmb{\sigma} T_1}=\epsilon I_{2\times2}+i\mathbf{r}\cdot\pmb{\sigma}\nonumber\\&&\equiv e^{-iH_\text{eff}T},\label{utt}\\
\epsilon&=&\cos(E_1T_1)\cos(E_2T_2)-\underline{\mathbf{h}}_1\cdot\underline{\mathbf{h}}_2\sin(E_2T_2)\nonumber\\&&\times\sin(E_1T_1),~~~\label{epsl}\\
 \mathbf{r}&=& \underline{\mathbf{h}}_1\times\underline{\mathbf{h}}_2\sin(T_1E_1)\sin(T_2E_2)-\underline{\mathbf{h}}_2\cos(T_1E_1)\nonumber\\ &&\times\sin(T_2E_2)-\underline{\mathbf{h}}_1\cos(T_2{E}_2)\sin(T_1{E}_1),\label{varepsilonds}
\end{eqnarray}
where and $T=T_1+T_2$.
Using the Euler's formula again, we can infer $H_\text{eff}$ from Eq. \eqref{utt} as
\begin{equation}
H_\text{eff}=-\arccos(\epsilon)\mathbf{r}\cdot\pmb{\sigma}/[\sin(\arccos(\epsilon))T]
\end{equation}
Due to $\epsilon^2+\mathbf{r}\cdot \mathbf{r}=1$, $H_\text{eff}$ can be simplified into
\begin{equation}
H_\text{eff}=\arccos(\epsilon)\mathbf{r}\cdot\pmb{\sigma}/[\sqrt{1-\epsilon^2}T]=\arccos(\epsilon)\underline{\mathbf{r}}\cdot\pmb{\sigma}/T.
\end{equation}
The quasienergy bands touch at $0$ and $\pm\pi/T$, which occurs when $\epsilon=+1$ and $-1$, respectively. Therefore, we
obtain that the bands close from Eq. \eqref{epsl} when
\begin{eqnarray}
&&T_jE_j=n_j\pi,~n_j\in \mathbb{Z}, \label{gen}\\
&&\text{or}~
\begin{cases}
\underline{\mathbf{h}}_1\cdot\underline{\mathbf{h}}_2=\pm1\\
T_1{E}_1\pm T_2{E}_2=n\pi,~n\in\mathbb{Z}
\end{cases}\label{hh1}
\end{eqnarray}
at the quasienergy zero (or $\pi/T$) if $n$ is even (or odd).

\section{Chiral symmetry in momentum space}
For a two-band non-Hermitian system, the coefficient matrix of its Hamiltonian in operator basis can be written in the momentum space as $\mathcal{H}(k)=\mathbf{h}(k)\cdot\pmb{\sigma}$, where $\pmb{\sigma}$ are the Pauli matrices. If the parameters in $\mathbf{h}$ are periodically driven between two specific $\mathbf{h}_1$ and $\mathbf{h}_2$ within the respective time duration $T_1$ and $T_2$, then we, according to the Floquet theorem, can obtain the effective Hamlitonian as $\mathcal{H}_\text{eff}(k)\equiv \mathbf{h}_\text{eff}\cdot\pmb{\sigma}=i\ln[e^{-i\mathcal{H}_2(k)T_2}e^{-i\mathcal{H}_1(k)T_1}]$ with the Bloch vector $\mathbf{h}_\text{eff}({\bf k})=-\arccos (\epsilon) \underline{\mathbf{r}}/T$ and
\begin{eqnarray}
\epsilon&=&\cos(T_1 E_1)\cos(T_2 E_2)-\underline{\mathbf{h}}_1\cdot\underline{\mathbf{h}}_2\sin(T_1{E}_1)\sin(T_2E_2),\nonumber\\
\mathbf{r}&=& \underline{\mathbf{h}}_1\times\underline{\mathbf{h}}_2\sin(T_1E_1)\sin(T_2E_2)-\underline{\mathbf{h}}_2\cos(T_1E_1)\nonumber\\ &&\times\sin(T_2E_2)-\underline{\mathbf{h}}_1\cos(T_2{E}_2)\sin(T_1{E}_1),\label{varepsilon}
\end{eqnarray}
where $T=T_1+T_2$, $\underline{\mathbf{h}}_j=\mathbf{h}_j/E_j$, and $E_j=\sqrt{\mathbf{h}_j\cdot\mathbf{h}_j}$ is the complex eigen energies of $\mathcal{H}_j({\bf k})$. We can see that $\mathcal{H}_\text{eff}(k)$ does not inherit the chiral symmetry of $\mathcal{H}_j(k)$ due to the presence of the first term of Eq. \eqref{varepsilon}.

We propose the following scheme to resolve this problem. A similarity transformation $G_1=e^{-i\mathcal{H}_1(k)T_1/2}$ converts the evolution operator $U_{T}$ to $\tilde{U}_{T,1}= {U}'_1 {U}'_2$ with ${U}'_1=e^{-i\mathcal{H}_1(k)T_1/2}e^{-i\mathcal{H}_2(k)T_2/2}$ and $ {U}'_2=e^{-i\mathcal{H}_2(k) T_2/2}e^{-i\mathcal{H}_1(k) T_1/2}$. According to Eq. \eqref{varepsilon}, we have ${U}'_j=\epsilon'_j I_{2\times 2}+i\mathbf{r}'_j\cdot\pmb{\sigma}$ with $\epsilon'_1=\epsilon'_2$ and
$\mathbf{r}'_j=(-1)^j a \underline{\mathbf{h}}_1\times\underline{\mathbf{h}}_2-b\underline{\mathbf{h}}_2-c\underline{\mathbf{h}}_1$,
where $a=\sin (T_1 E_1/2)\sin (T_2E_2/2)$, $b=\cos (T_1E_1/2)\sin (T_2E_2/2)$, and $c=\cos (T_2E_2/2)\sin (T_1E_1/2)$. Then we can obtain
$ \tilde{U}_{T,1}=\epsilon' I_{2\times 2}+i\mathbf{r}'\cdot\pmb{\sigma}$ with $\epsilon'=(\epsilon'_1 )^2-\mathbf{r}'_1\cdot\mathbf{r}'_2$ and
\begin{eqnarray}
\mathbf{r}'&=&2\underline{\mathbf{h}}_1(-\epsilon'_1 c+ac\underline{\mathbf{h}}_1\cdot\underline{\mathbf{h}}_2+ab)\nonumber\\
&&-2\underline{\mathbf{h}}_2(\epsilon'_1 b+ac+ab\underline{\mathbf{h}}_1\cdot\underline{\mathbf{h}}_2).\label{r'}
\end{eqnarray}
Equation~\eqref{r'} implies that if $\mathcal{H}_1(k)$ and $\mathcal{H}_2(k)$ have the chiral symmetry with a common symmetry operator, then $\tilde{U}'_{T,1}$ would inherit this symmetry. The similar result can be obtained by $ {G}_2=e^{i \mathcal{H}_2(k)T_2/2}$, which converts ${U}_T$ into $\tilde{U}_{T,2}= {U}'_2 {U}'_1$. Leaving the quasienergy spectrum unchanged, the similarity transformations $G_j$ have succeeded in making $\tilde{\mathcal{H}}_{\text{eff},j}(k)\equiv {i\over T}\ln \tilde{U}_{T,j}$ preserve the chiral symmetry in $\mathcal{H}_j(k)$.

\section{Generalized Brillouin zone}
According to Ref. \cite{PhysRevLett.121.086803}, the skin effect in the static non-Hermitian SSH model can be curled by a similarity transformation, which reads $\mathcal{S}=\text{diag}(1,r,r,r^2,\cdots,r^{L-1},r^L)$ in the real-space sublattice basis. It converts Eq. (6) into $\bar{H}=\mathcal{S}^{-1}H\mathcal{S}$ with
\begin{equation}
\bar{H}=\sum_{l=1}^L[r(t_1+{\gamma\over 2}) a^{\dag}_lb_l+r^{-1}(t_1-{\gamma\over 2})b^{\dag}_l a_l+t_2(a^{\dag}_lb_{l-1}+\text{h.c.})].\label{Hmats2}
\end{equation}One can readily check that, as long as $r=\sqrt{\lvert (t_1-\gamma/2)/(t_1+\gamma/2) \lvert}$, all the bulk states of $\bar{H}$ do not reside in the edges anymore and thus the skin effect is curled. Comparing the forms of $H$ and $\bar{H}$ in the momentum space, we see that the usual BZ $e^{ik}$ in $H$ is replaced by $re^{ik}\equiv\beta$ in $\bar{H}$. Thus $\beta$ defines a generalized BZ.

An important observation is that the similarity transformation $\mathcal{S}$ has nothing effect on the $t_2$ term. When the periodic driving \begin{equation}
t_2(t) =\begin{cases}f ,& t\in\lbrack mT, mT+T_1)\\q\,f,& t\in\lbrack mT+T_1, (m+1)T), \end{cases}~m\in \mathbb{Z}, \label{procotol}
\end{equation} is applied on $t_2$, we readily have
\begin{equation}
\bar{U}_T=\mathcal{S}^{-1}U_T\mathcal{S}=e^{-i\bar{H}_2T_2}e^{-i\bar{H}_1T_1},
\end{equation}where $\bar{H}_j$ are Eq. \eqref{Hmats2} with $t_2=f$ and $qf$, respectively. Since no skin effect is present in $\bar{H}_j$, neither to $H_\text{eff}={i\over T}\ln U_T$. We have successfully remove the skin effect by the same generalized BZ defined by $\mathcal{S}$ as the static case.

On the other hand, if the periodic driving is applied on $t_1$, the generalized BZ would be changed. One can resort to the method in Refs. \cite{PhysRevLett.123.066404,yang2019auxiliary} to calculate the generalized BZ in the general case.
\section{$Z_2$ topological invariant}
In the widely used application of Floquet engineering, there exist some periodic driving cases in which even the proposed similarity transformations cannot restore the chiral symmetry. A typical example is the three-step periodic driving to create the discrete time crystal \cite{Zhang2017}. If the chiral symmetry cannot be recovered, the periodically driven system intrinsically belongs to the D topological class, where the winding number is ill-defined \cite{RevModPhys.88.035005}. We can define a new $Z_2$ topological invariant in the real space to characterize this kind of non-Hermitian Floquet topological phases. Inspired by the definition of electronic polarization of the two-dimensional Hermitian system in the static case \cite{PhysRevB.100.245135}, we can construct an open-boundary bulk-band $Z_2$ topological invariant $\nu=2P$ for our one-dimensional periodically driven non-Hermitian system from the electronic polarization
\begin{equation}
P=\Big[\frac{1}{2\pi}\text{Im}\ln\det\mathcal{U}-\sum_{l,l',s,s'} \frac{X_{ls,l's'}}{2L}\Big]\mod 1,
\end{equation}
where the elements of $\mathcal{U}$ read $\mathcal{U}_{mn}\equiv \langle m^\text{L}|e^{i2\pi X/L}|n^\text{R}\rangle$ with ${\mathcal H}_{\text{eff}}|n^\text{R} \rangle=\varepsilon_{n}|n^\text{R}\rangle$, ${\mathcal H}^{\dag}_{\text{eff}}|n^\text{L} \rangle=\varepsilon^{*}_{n}|n^\text{L}\rangle$, and $X$ is the coordinate operator, namely $X_{ls,l's'}=l\delta_{ll'}\delta_{ss'}$ with $s,s'=A,~B$ being the sublattices. The $Z_2$ topological invariant $\nu$ can only distinguishes between even or odd number of pairs of edge states.

The one-period evolution operator for a three-step periodic driving is given by \cite{Zhang2017}
\begin{equation}
U_T=e^{-i\mathcal{H}_3({\bf k})T_3}e^{-i\mathcal{H}_2({\bf k})T_2}e^{-i\mathcal{H}_1({\bf k})T_1}.
\end{equation}
In the similar manner as the two-step periodic driving in the main text, we can derive that the phase transition occurs for the $k$ and driving parameters satisfying
\begin{eqnarray}
&&T_jE_j=n_j\pi,~n_j\in \mathbb{Z}, \label{gen}\\
&&\text{or}~
\begin{cases}
\underline{\mathbf{h}}_1\cdot\underline{\mathbf{h}}_2=\pm1,\,\, \underline{\mathbf{h}}_1\cdot\underline{\mathbf{h}}_3=\pm1,\\
T_1{E}_1\pm T_2{E}_2\pm T_3E_3=n\pi,~n\in\mathbb{Z}
\end{cases}\label{hh1}
\end{eqnarray}
at the quasienergy zero (or $\pi/T$) if $n$ is even (or odd).

Taking the non-Hermitian Su-Schrieffer-Heeger model as an example, we illustrate the performance of the $Z_2$ topological invariant $\nu$ in characterizing the Floquet topological phases. After introducing the generalized Brillouin zone via replacing $e^{ik}$ by $\beta=\sqrt{\frac{t_1-\gamma/2}{t_1+\gamma/2}}e^{ik}$, the non-Bloch Hamiltonian with ${\bf h}=[t_1+t_2(\beta+\beta^{-1})/2,i[\gamma+t_2(\beta^{-1}-\beta)]/2,0]$ is obtained \cite{PhysRevLett.121.086803}.
We consider an experimentally accessible periodic-driving protocol
\begin{equation}
t_2(t) =\begin{cases}q_1\,f ,& t\in\lbrack mT, mT+T_1)\\q_2\,f,& t\in\lbrack mT+T_1, mT+T_1+T_2)\\q_3\,f ,& t\in \lbrack  mT+T_1+T_2,(m+1)T)  \end{cases}.
\end{equation}
Next, we derive the condition for the band closing. To satisfy the first line of Eq. \eqref{hh1}, $k$ in $\beta$ can only be $\alpha=0$ or $\pi$, at which the Hamiltonians read
\begin{equation}
\mathcal{H}_j|_{k=\alpha}=\frac{e^{i\alpha}q_j f+\kappa}{\kappa}(t_1\sigma_x+\frac{i\gamma}{2}\sigma_y).
\end{equation}
with $\kappa=\sqrt{t_1^2-\gamma^2/4}$. Due to $[ \mathcal{H}_i, \mathcal{H}_j]=0$, we obtain the effective Hamiltonian at $k=\alpha$
     \begin{equation}
      \mathcal{H}_\text{eff}|_{k=\alpha}=(\mathcal{H}_3T_3+\mathcal{H}_2T_2+\mathcal{H}_1T_1)/T.
     \end{equation}
Then the quasienergy bands close at the quasienergies $0$ and $\pi/T$ when
     \begin{equation}
     \sum_{j=1}^3(\kappa+e^{i\alpha}q_jf)T_j=n_\alpha\pi\label{scond}
     \end{equation}
for $n_\alpha$ being even and odd numbers, respectively.
This determines the critical points for topological phase transition. Here we have assumed that the parameters satisfy $t_1>\gamma/2>0$.

\begin{figure}[tbp]\centering
\includegraphics[width=\columnwidth]{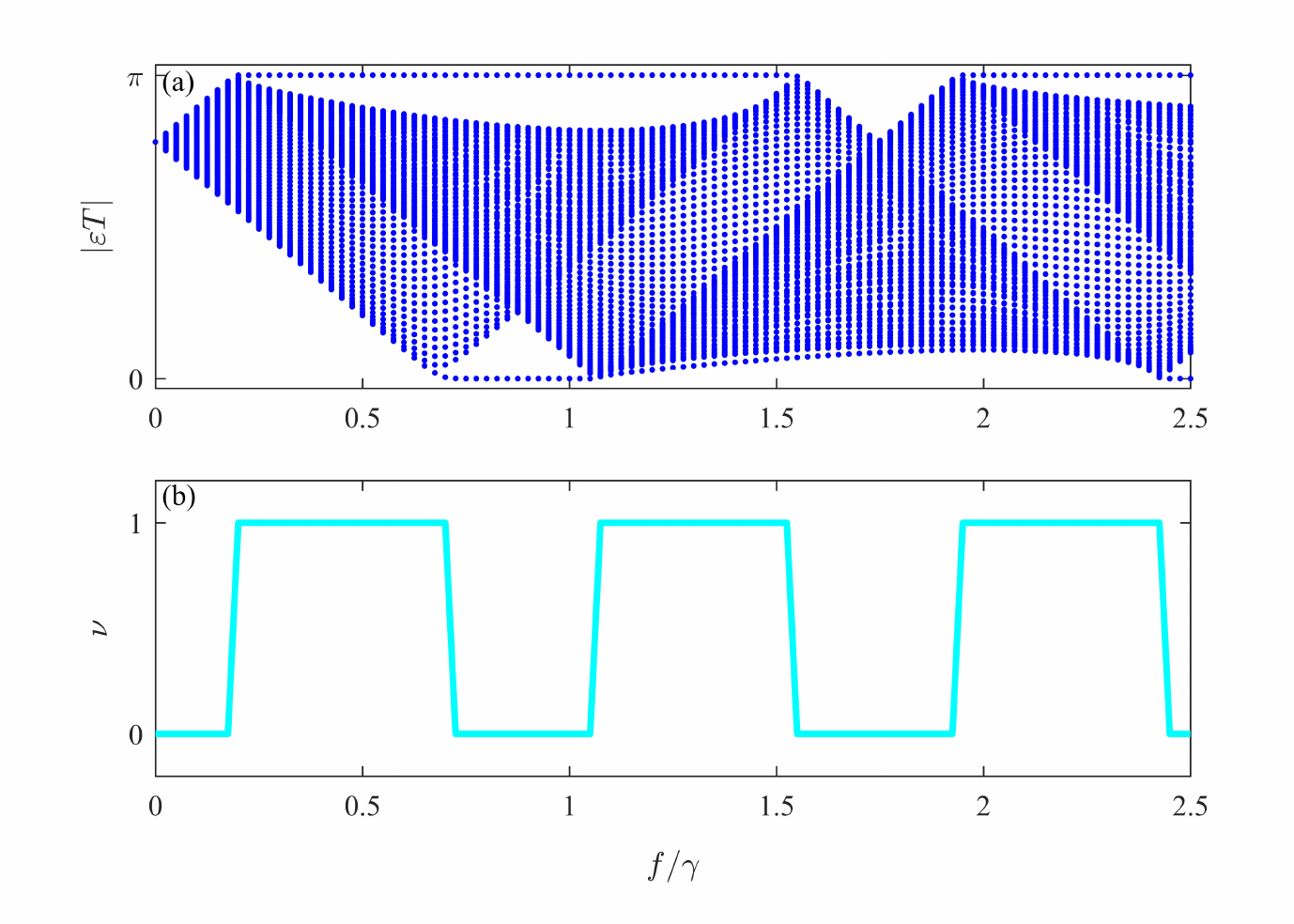}
\caption{(a) Quasienergy spectra with the change of the driving amplitude under the open-boundary condition. (b) the $Z_2$ topological invariant $\nu$. We use $t_1=1.45\gamma$, $T_1=T_2=T_3=0.6\gamma^{-1}$, $q_1=1.0$, $q_2=2$, $q_3=3$, and $L=80$.}\label{Sengsp}
\end{figure}
Figure \ref{Sengsp}(a) shows the quasienergy spectrum under the open-boundary condition. Different from the static case, the extra $\pi/T$-mode edge states is induced by the periodic driving. The $\pi/T$-mode band-closing points at $f=0.19\gamma$, $1.55\gamma$, and $1.94\gamma$ can be obtained from Eq. \eqref{scond} with $n_\alpha=1_0$, $-1_\pi$, and $3_0$, respectively. The 0-mode ones at $f=0.68\gamma$, $1.06\gamma$, and $2.43\gamma$ is explainable from Eq. \eqref{scond} with $n_\alpha=0_\pi$, $2_0$, and $-2_\pi$, respectively. The $Z_2$ topological invariant $\nu$ depicted in Fig. \ref{Sengsp}(b) perfectly describes the Floquet topological phases in our periodically driven non-Hermitian system. If $\nu=1$, then the system holds an odd number of pairs of edge states. If $\nu=0$, then the system has an even number of pairs of edge states.
\bibliography{references.bib}